\RequirePackage{fix-cm}
\documentclass[smallextended]{svjour3}       
\smartqed  
\usepackage{CJKutf8}
\usepackage{graphicx}
\usepackage{amsmath,amssymb,amsfonts}
\usepackage{algorithmic}
\usepackage{enumitem}
\usepackage{url}
\usepackage{color}
\usepackage{subfigure}
\usepackage{adjustbox}
\usepackage{array, boldline, makecell, booktabs}
\usepackage[mathscr]{euscript}
\usepackage{lscape}
\usepackage{tabularx}
\usepackage{multirow}
\usepackage{xcolor}
\usepackage{makecell}

\begin{document}

\title{\textsc{CHECKED}: Chinese COVID-19 Fake News Dataset 
}


\author{Chen Yang       \and
        Xinyi Zhou      \and
        Reza Zafarani
}


\institute{Chen Yang, Xinyi Zhou, Reza Zafarani \at
              Data Lab, Department of Electrical Engineering and Computer Science, Syracuse University\\
              \email{cyang03@syr.edu, zhouxinyi@data.syr.edu, reza@data.syr.edu}           
}

\date{Received: date / Accepted: date}

\maketitle

\begin{abstract}
COVID-19 has impacted all lives. To maintain social distancing and avoiding exposure, works and lives have gradually moved online. Under this trend, social media usage to obtain COVID-19 news has increased. Also, misinformation on COVID-19 is frequently spread on social media. In this work, we develop \textsf{CHECKED}, the first Chinese dataset on COVID-19 misinformation. \textsf{CHECKED} provides a total 2,104 verified microblogs related to COVID-19 from December 2019 to August 2020, identified by using a specific list of keywords. Correspondingly, \textsf{CHECKED} includes 1,868,175 reposts, 1,185,702 comments, and 56,852,736 likes that reveal how these verified microblogs are spread and reacted on Weibo.
The dataset contains a rich set of multimedia information for each microblog including ground-truth label, textual, visual, temporal, and network information. Extensive experiments have been conducted to analyze \textsf{CHECKED} data and to provide benchmark results for well-established methods when predicting fake news using \textsf{CHECKED}.
We hope that \textsf{CHECKED} can facilitate studies that target misinformation on coronavirus. The dataset is available at \url{https://github.com/cyang03/CHECKED}.

\end{abstract}



\vspace{-2mm}
\section{Introduction}
\label{sec:intro}

Starting from its first case, confirmed on December 31 in Wuhan, the novel coronavirus has surged into a world phenomenon rapidly. On January 30, the World Health Organization (WHO) has declared its outbreak as a global emergency~\cite{sohrabi2020world}. 
As of October 13, the COVID-19 outbreak has caused over 3.7 million confirmed cases and over 1 million deaths worldwide.\footnote{\url{https://covid19.who.int/}\label{ft:who}} To combat the epidemic, maintaining social distance has been considered effective. In turn, working and studying from home has become a new trend. With the decrease in physical social contacts and the rise of anxiety on the pandemic, the frequency of social media usage has increased. The COVID-19 outbreak as an international public health emergency is closely connected with individuals' health and lives. Any news or information about the COVID-19 or a potential cure highly attracts public attention and influences social media. Therefore, it is of crucial importance to ensure information spread on COVID-19 is credible.

With more information on COVID-19, people gain a deeper understanding. To that end, a number of COVID-related datasets have been released and studied. Existing datasets have contributed to collecting either (i) Chinese COVID data without identifying news credibility (e.g., see \cite{hu2020weibo,gao2020naist}); or (ii) non-Chinese COVID data for news credibility (e.g., see \cite{zhou2020recovery,cui2020coaid,li2020mmcovid}). Therefore, we are motivated to build a dataset which contains data from Chinese social media and includes ground-truth labels (i.e., true/false).

Weibo (\url{weibo.com}), as a platform for information sharing, dissemination, and acquisition based on user relations, is one of the most popular social media in China. According to Weibo's first-quarter earnings report for 2020,\footnote{\url{http://ir.weibo.com/news-releases/news-release-details/weibo-reports-first-quarter-2020-unaudited-financial-results}} Weibo passed 500 million monthly active users and 200 million daily active users in March. Due to the large number of active users, we consider Weibo as one of the most used Chinese social media for people to access information related to COVID-19. Furthermore, Weibo's Community Management Center\footnote{\url{https://service.account.weibo.com/} (sign in required)\label{footnote:weibo_management_center}}, which timely provides fact-check results of suspected news information (microblogs) verified by experts, has been widely accepted for researching rumors and fake news~\cite{jin2017multimodal,wang2018eann}. Therefore, we consider Weibo as a reliable source of collecting verified microblogs.

We develop \textsf{CHECKED}, the first Chinese COVID-19 social media dataset. \textsf{CHECKED} contains 344 ``Fake'' microblogs and 1760 ``Real'' microblogs from December 2019 to August 2020, along with 1,868,175 reposts and 1,185,702 comments that reveal how these verified microblogs were spread and reacted on Weibo. The main contributions of this work are: 
\begin{enumerate}
    \item We introduce the first fact-checked Chinese COVID-19 social media dataset, which enables more research on tracing the spread of microblogs misinformation and on analyzing content patterns in COVID-19 fake news.
    
    \item We contribute the dataset with a rich set of features on microblogs related to COVID-19. We collect textual, visual, and video information as well as network and temporal information of a microblog.
    
    \item We conduct comprehensive experiments to analyze \textsf{CHECKED} data. We provide benchmark results for well-established methods when identifying fake news using this data. Data and codes are all public (see \url{https://github.com/cyang03/CHECKED}).
\end{enumerate} 

The rest of the paper is organized as follows. Literature review is first conducted in Section~\ref{sec:review}. We explain how we collect data in Section~\ref{sec:data_collection}. Data are then analyzed in Section~\ref{sec:analysis}, followed by benchmark results for well-established methods when applied to \textsc{CHECKED} in Section \ref{sec:baseline}. Finally, we conclude in Section~\ref{sec:conclusion}.

\vspace{-2mm}
\section{Related Work}
\label{sec:review}

Our related work can be organized into (I) social media datasets on COVID-19; (II) COVID-19 datasets for news credibility; and (III) Weibo data for news credibility research.

\paragraph{I. COVID-19 Social Media Datasets}
Social media can provide a wealth of information, especially during the pandemic. Thus, a number of social media datasets on COVID-19 have emerged. Chen et al.~\cite{chen2020tracking} released the first COVID-19 dataset collected from Twitter, tracking the information related to coronavirus from January 2020 till the present with continuous updates. There are a few social media COVID-19 datasets in Chinese (e.g., Weibo-COV~\cite{hu2020weibo} and NAIST-COVID~\cite{gao2020naist}). Weibo-COV~\cite{hu2020weibo} is a large-scale COVID-19 dataset with 40 million microblogs in Chinese from December 2019 to April 2020. The dataset provides textual information, geographical information, and response information. NAIST-COVID~\cite{gao2020naist} is a large-scale multilingual COVID-19 dataset which consists of English (16 million), Japanese (9 million), and Chinese (180 thousand) microblogs from Twitter and Weibo from January 20 to March 24.
While these datasets are large, they do not provide visual information or labels on news credibility.

\paragraph{II. COVID-19 News Credibility Datasets}
With the spread of COVID-19, rumors and fake news related to it have also spread. Thus, constructing a COVID-19 dataset with labels on news credibility, which often relies on verification by domain experts, is invaluable for research. ReCOVery dataset~\cite{zhou2020recovery} contains over 2,000 verified news articles on COVID-19 from extremely reliable and unreliable outlets. Both textual and visual information of news articles are collected, along with over 140,000 tweets by tracking news URLs on Twitter.
CoAID dataset~\cite{cui2020coaid} includes verified 3,252 news articles and claims and 851 microblogs on Twitter about COVID-19, which correspond to 296 thousands related user engagements from December 2019 to July 2020. 
MM-COVID dataset~\cite{li2020mmcovid} provides multilingual fact-checked news statements in six languages (English, Spanish, Portuguese, Hindi, French and Italian) and the relevant social context. 
The Spanish dataset in MM-COVID dataset is the largest non-English COVID-19 dataset, containing 3,213 verified news articles and 28,824 related user engagements.
Nevertheless, few Chinese COVID-19 dataset has been constructed to support Chinese COVID-19 news credibility research, which motivates our work.

\paragraph{III. Weibo Data for News Credibility Research}
Weibo's Community Management Center\textsuperscript{\ref{footnote:weibo_management_center}}, which timely provides fact-check results of suspected news information (microblogs) verified by experts, has been widely accepted for researching rumors and fake news. For example, the dataset constructed in \cite{jin2017multimodal} and utilized in \cite{wang2018eann} contains 4,749 fake news, collected from Weibo's Community Management Center, and 4,779 real news, collected from Xinhua News Agency.
Though these Weibo data have greatly contributed to fake news and rumor research, they were collected before COVID-19 pandemic and would not be timely updated; hence can be hardly used, in particular, to combat COVID-19 infodemic. 

\vspace{1em}
Compared to the datasets mentioned above, \textsf{CHECKED} is the first Chinese COVID-19 dataset with ground truth labels. \textsf{CHECKED} includes both textual and visual information, as well as the label of news credibility and propagation information in terms of comments and reposts. \textsf{CHECKED} is comparable in to most non-English COVID-19 datasets for news credibility research.

\vspace{-2mm}
\section{Data Collection}
\label{sec:data_collection}

We detail the data collection method which answers the following questions: (I) Where can we obtain the ground-truth labels (\textit{real} or \textit{fake}) on news events?; (II) How can we identify whether a news event is relevant to the coronavirus or not?; (III) What time intervals should be considered to achieve an extensive yet efficient search coverage?; and (IV) How can the information on news events (meta-data and its spread on Weibo) be tracked and stored in the dataset.

\paragraph{I-1 Collecting Fake News.}
Weibo provides the Weibo Community Management Center,\textsuperscript{\ref{footnote:weibo_management_center}} an official service where users can report either a microblog that (i) contains false information, (ii) releases user privacy without permission, (iii) has evidence of cyberbullying, and (iv) shows plagiarism; or belongs to a user who (v) impersonates as someone else. Experts in charge are then involved in verifying and ultimately, share their detailed evaluation of the reports at the platform publicly. This official service has helped verify over two million microblogs and remove tens of thousands of mis/disinformation.
Our fake news collection focuses on these microblogs, which having been reported and labeled as false information in the Weibo Community Management Center.

\paragraph{I-2 Collecting Real News.}
To collect credible COVID-19 microblogs, we rely on two official reports, the (I) ``2019 White Paper on the Social Value of Chinese Online Medium''\footnote{\url{https://www.ndrc.gov.cn/xxgk/jd/wsdwhfz/202004/P020200414717451252380.pdf} (Chinese)} and the (II) ``Research Report on the Public Awareness and Information Dissemination of COVID-19''\footnote{\url{http://www.sic.gov.cn/archiver/SIC/UpFile/Files/Default/20200226101829580669.pdf} (Chinese)}, provided by the State Information Center in (Administration Center of China E-government Network), the public institution directly affiliated to the National Development and Reform Commission.
Both reports provide rankings based on different criteria for Chinese online media. Specifically,
\begin{itemize}
    \item In the white paper, domain experts evaluate 24 Chinese major online media based on eight primary criteria and 28 secondary criteria. The evaluation covers aspects of the
    (i) quality and diversity in platform construction;
    (ii) social influence including the platform popularity;
    (iii) activity; 
    (v) reputation in the field and among online users; and (vi) how the medium contributes to charity.
    
    \item The COVID-19 research report ranks the performance of online media during the pandemic. The ranking data is obtained through $\sim$3,000 valid surveys taken by online users, which focuses on the platform content trustworthiness, communication capacity, and social responsibility.
\end{itemize} 
We select the Weibo account, People's Daily,\footnote{\url{https://weibo.com/rmrb}} to collect real news. As China's largest newspaper group, People's Daily is ranked first in both reports, with over 120 million followers and over 120,000 microblogs on Weibo.

\begin{CJK*}{UTF8}{gbsn}
\begin{table}[t]
\centering
\caption{List of Keywords Relevant to COVID-19}
\label{tab:keywords}
\begin{tabular}{cll}
\toprule[0.8pt]
\textbf{Categories} & \textbf{Keywords} & \textbf{English Translation} \\ \midrule[0.5pt]
\multirow{6}{*}{\makecell{Coronavirus \\ and COVID-19}} & 冠状病毒 & Coronavirus \\  
 & 新冠肺炎 & COVID-19 \\  
 & 新冠 & (Abbr.) Coronavirus/COVID-19 \\ 
 & Coronavirus & N/A \\ 
 & SARS-CoV-2 & N/A \\ 
 & COVID & N/A \\ \hline
\multirow{6}{*}{Pandemic} & 疫情 & Pandemic/epidemic \\  
 & 疫区 & Pandemic/epidemic area \\  
 & 传染, 感染 & Infection \\  
 & 确诊 & Confirmed case \\  
 & 死亡病例 & Death case \\  
 & 输入病例, 输入性传播 & Imported case \\ \hline
\multirow{8}{*}{\makecell{Figures and \\ organizations}} & 世界卫生组织 & WHO \\  
 & 世卫 & (Abbr.) WHO \\  
 & 钟南山 & Nanshan Zhong \\  
 & 张文宏 & Wenhong Zhang \\  
 & 李文亮 & Wenliang Li \\  
 & 福奇 & Fauci \\ 
 & WHO & N/A \\ 
 & CDC & N/A \\ \hline
\multirow{8}{*}{\makecell{Medical\\ supplies}} & 试剂盒 & Testing kit \\  
 & 核酸检测 & Nucleic Acid Test \\  
 & 疫苗 & Vaccine \\  
 & 抗体 & Antibody \\  
 & 火神山 & Huoshenshan \\  
 & 雷神山 & Leishenshan \\ 
 & 口罩 & Mask \\ 
 & N95 & N/A \\ \hline
\multirow{7}{*}{Policies} & 隔离 & Quarantine \\
 & 封城 & Lockdown \\  
 & 防控 & Prevention And Control \\  
 & 群体免疫 & Herd (community) immunity \\  
 & 健康码, 健康宝 & Health code \\  
 & 战疫, 抗疫 & Combat COVID-19 \\  
 & 援鄂 & Love for Wuhan \\
 \bottomrule[0.8pt]
 \multicolumn{3}{l}{English keywords are case-insensitive.} 
\end{tabular}
\end{table} 
\end{CJK*}

\paragraph{II. Identifying Microblog Topics.}
After a broad investigation, we finally select a total of 39 keywords to determine whether a microblog is relevant to COVID-19 or not. All the identified keywords are listed in Table \ref{tab:keywords}, which relate to (i) the name of the virus and disease, (ii) pandemic, (iii) the figures and organizations playing a pivotal role in combating the pandemic, (iv) medical supplies, and (v) policies. Most keywords are in Chinese (32 keywords) and a few, as commonly-used professional terms (e.g., N95), are in English (6 keywords). English keywords are case-insensitive. A microblog containing \textit{any} of the keywords in our list is regarded as relevant to COVID-19. To guarantee the data quality and news relevance, we also manually checked all fetched news articles and removed those irrelevant to COVID-19.

\paragraph{III. Collection Interval.}
For an unbiased data collection, we collected both real and fake microblogs (news events) from December 2019, when the coronavirus was first identified in Wuhan, Heibei, China~\cite{huang2020clinical}, to August 2020 (nine months).

\paragraph{IV. Collecting News Events and Tracking their Spread on Weibo.} Given sources for real and fake news, keywords, and data collection time interval for coronavirus pandemic, we collect microblogs, each with the following components:
\begin{itemize}
\item $\mathtt{id}$: Each microblog is identified by a unique 16-digit ID number assigned by Weibo. To protect users' privacy, we have hashed each ID to 32 digits in the \textsf{CHECKED} dataset.
\item $\mathtt{label}$: $\mathtt{label}$ is either ``real'' or ``fake'', as the label of each microblog.
\item $\mathtt{analysis}$: It is an official report including detailed analysis and a result on detecting fake news from Weibo experts.
\item $\mathtt{date}$: $\mathtt{date}$ is the time  each microblog is posted in $\mathtt{yyyy}$-$\mathtt{mm}$-$\mathtt{dd}$  $\mathtt{hh}$:$\mathtt{mm}$ format. 
\item $\mathtt{user\_id}$: $\mathtt{user\_id}$ is a unique 10-digit ID number assigned to the user by Weibo. Each user can change his or her ID only once to a string formed by 4 to 20 characters, in which letters are allowed. To protect users' privacy, we have also hashed each user's ID to 32 digits in the \textsf{CHECKED} dataset.
\item $\mathtt{text}$: It contains the textual information of a microblog.
\item $\mathtt{pic\_url}$: the URL for the visual information of the microblog; Users are allowed to attach no more than 18 images to each microblog. Such visual information has been recently developed for multimodal fake news detection~\cite{zhou2020multimodal,wang2018eann,jin2017multimodal}
\item $\mathtt{video\_url}$: this URL for the video information of the microblog. Each microblog (i) can only include at most one video; and (ii) cannot contain both a video and an image.
\item $\mathtt{comment\_num}$, $\mathtt{repost\_num}$, and $\mathtt{like\_num}$: the numbers of comments, forwards, and likes for a microblog, respectively. 
\item $\mathtt{comments}$: detailed information on user comments for each microblog. The information includes the hashed ID, date, and content of comments (microblogs), and the hashed ID of commenters (users). For each comment, no more than one image and no video are allowed.
\item $\mathtt{reposts}$: detailed information on user forwards for each microblog: the hashed ID, date, content of forwards (microblogs), and the hashed ID of forwarders (users). Similar to comments, each forward has at most one image and no video information. If a user forwards a repost with an image, the $\mathtt{pic\_url}$ of the new forward will also include this image along with the original image.
\end{itemize}

\begin{figure}[t]
    \centering
    \subfigure[Fake Microblog]{
    \includegraphics[width=\textwidth]{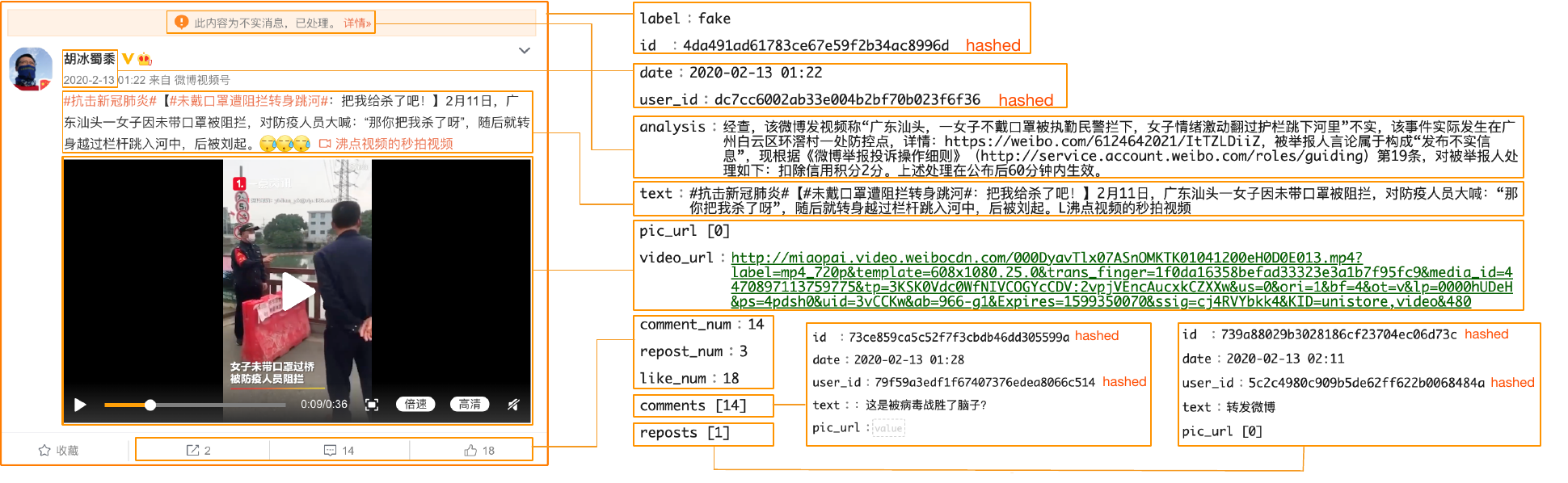}}
    \subfigure[Real Microblog]{
    \includegraphics[width=\textwidth]{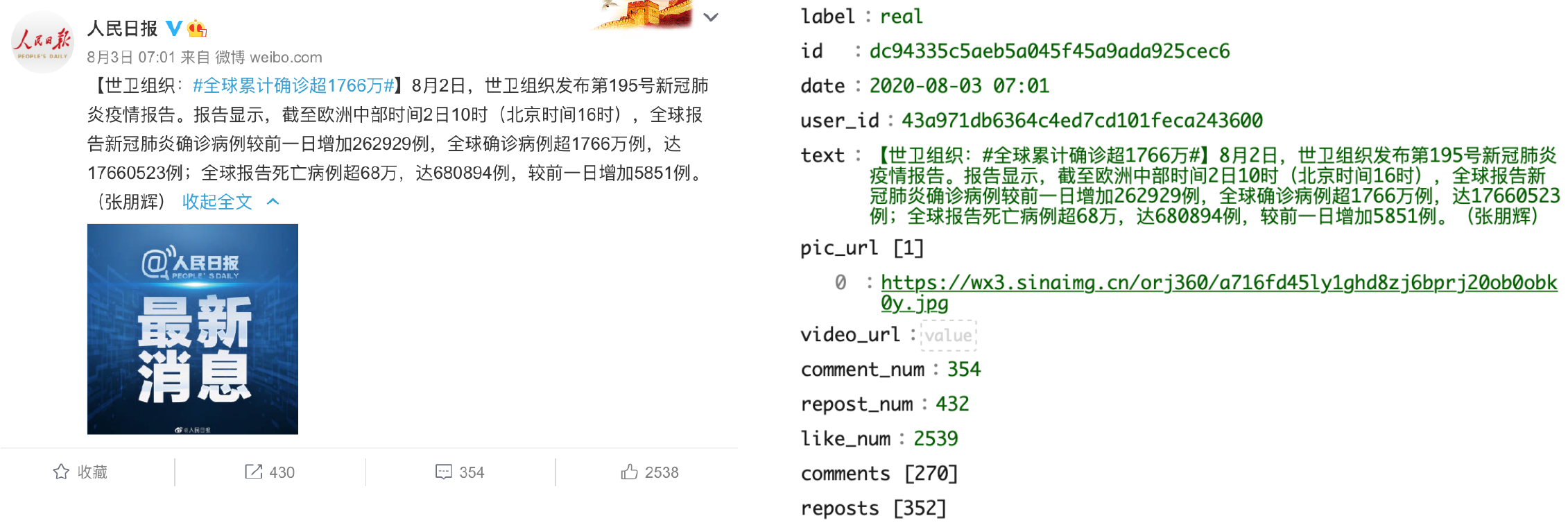}}
    \caption{Illustrations of Collected Microblogs}
    \label{fig:microblog_examples}
\end{figure}

We illustrate a real and fake microblog collected in our dataset in Figure \ref{fig:microblog_examples}. Meanwhile, we point out that, due to Weibo's access restriction and editable visibility,
it is possible that, for a few microblogs, their number of comments (forwards) in $\mathtt{comment}$ ($\mathtt{repost}$), which indicates the number visible on Weibo, is less than that of comments (forwards) in $\mathtt{comment\_num}$ ($\mathtt{repost\_num}$), which captures the actual number.

\vspace{-2mm}
\section{Data Analysis}
\label{sec:analysis}

The statistics of \textsf{CHECKED} dataset is presented in Table \ref{tab:statistics}. Next, we analyze the collected data in terms of (I) textual information, (II) visual information, (III) propagation and responses of collected microblogs.

\begin{table}[t]
\centering
\caption{Statistics of \textsf{CHECKED} Data}
\label{tab:statistics}
\begin{tabular}{lrrr}
\toprule[1pt]
 & \textbf{Real} & \textbf{Fake} & \textbf{All} \\ \midrule[0.5pt]
\textbf{\# Microblogs} & 1,760 & 344 & 2,104 \\
\multicolumn{1}{l}{\qquad with images} & 1,149 & 53 & 1,202 \\
\multicolumn{1}{l}{\qquad with video} & 563 & 106 & 669 \\
\multicolumn{1}{l}{\qquad with reposts} & 1,151 & 229 & 1,380 \\
\multicolumn{1}{l}{\qquad with comments} & 1,151 & 292 & 1,443 \\
\textbf{\# Reposts of microblogs} & 1,827,817 & 40,358 & 1,868,175 \\
\textbf{\# Comments of microblogs} & 1,169,246 & 16,456 & 1,185,702 \\
\textbf{\# Likes of microblogs} & 56,407,610 & 445,116 & 56,852,726 \\
\textbf{\# Weibo users} & 686,077 & 51,674 & 737,751 \\ \bottomrule[1pt]
\end{tabular} 
\end{table}

\begin{figure}[t]
    \begin{minipage}{\textwidth}
        \begin{minipage}[b]{0.81\textwidth}
        \includegraphics[width=\textwidth]{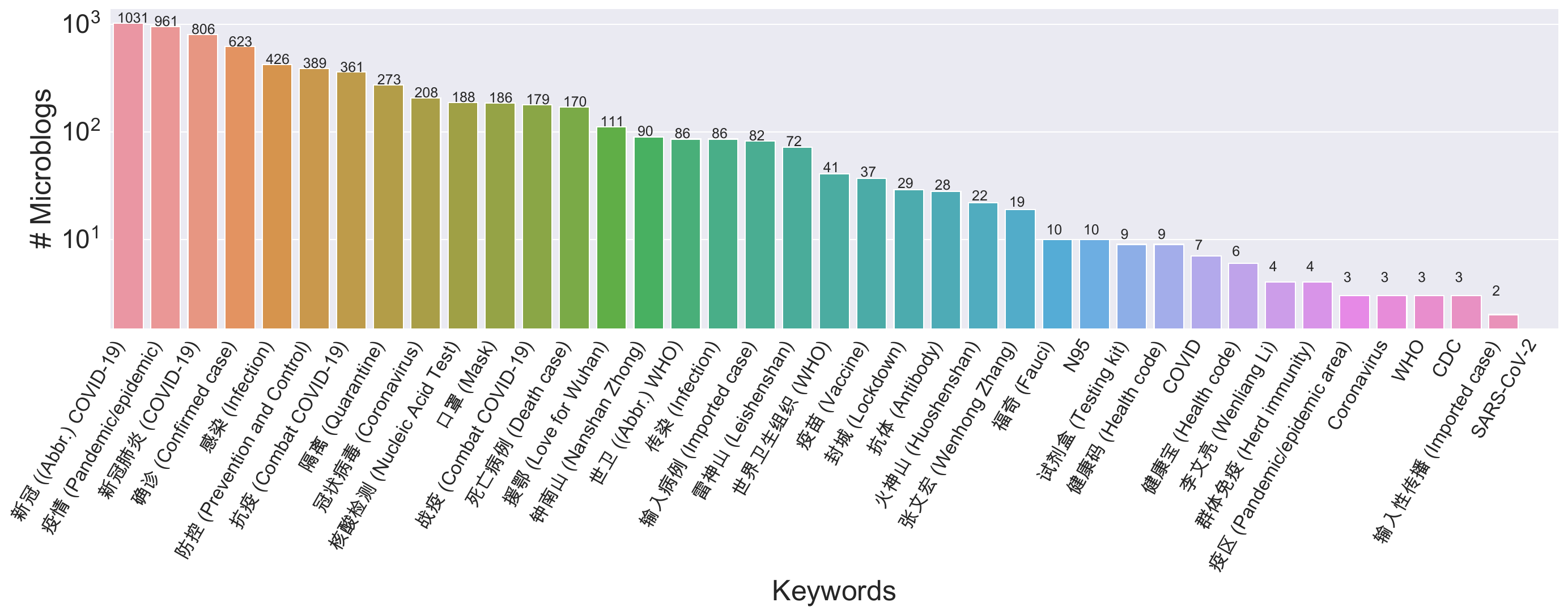}
        \caption{Distribution of Selected Keywords in Collected Microblogs}
        \label{fig:keyword_dist}
        \end{minipage}~
        \begin{minipage}[b]{0.18\textwidth}
        \includegraphics[width=\textwidth]{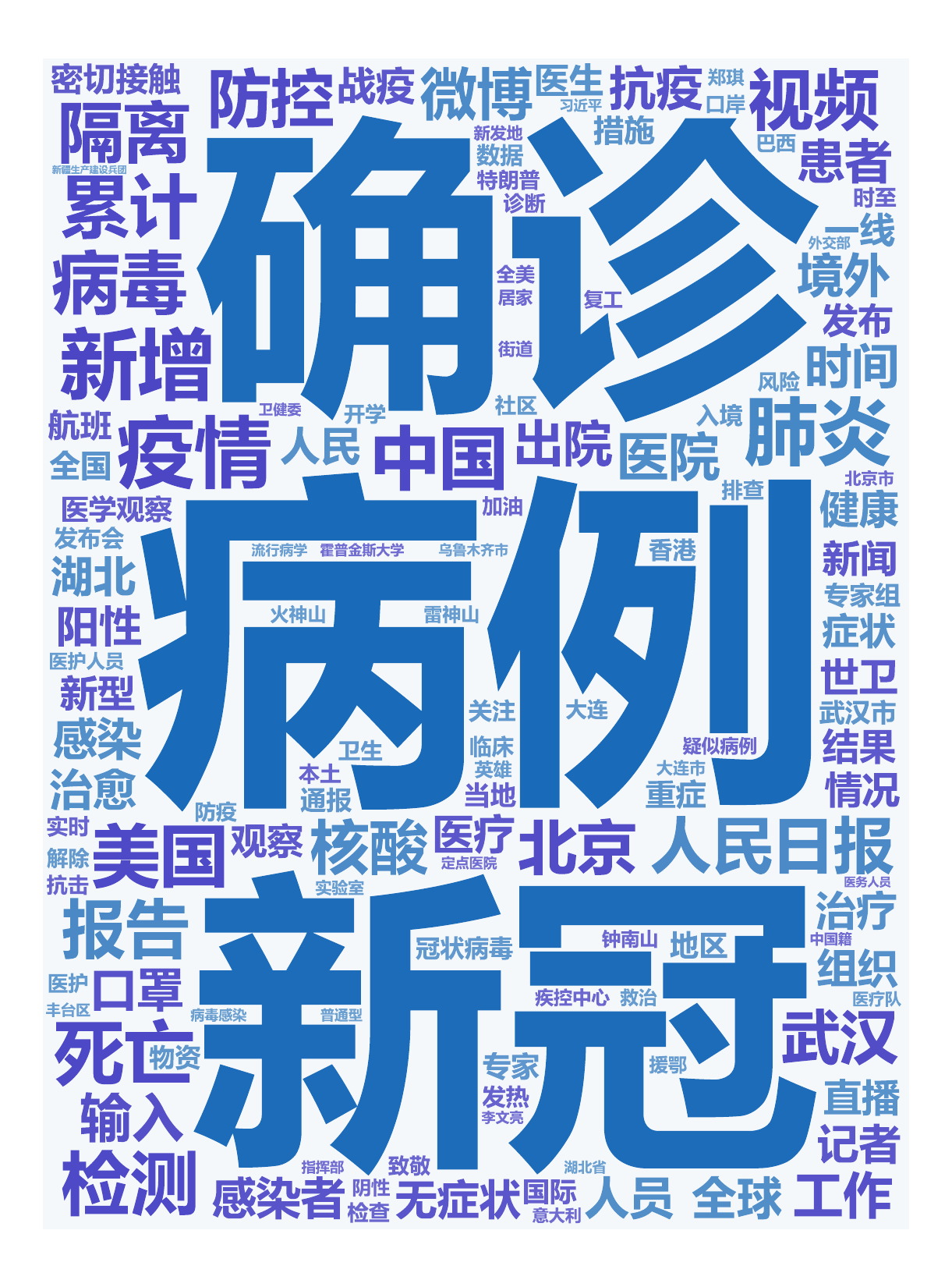}
        \vspace{1mm}
        \caption{Word Cloud}
        \label{fig:word_cloud}
        \end{minipage}        
    \end{minipage}
\end{figure}
\begin{figure}[t]
    \begin{minipage}{\textwidth}
        \begin{minipage}[b]{0.32\textwidth}
        \includegraphics[width=\textwidth]{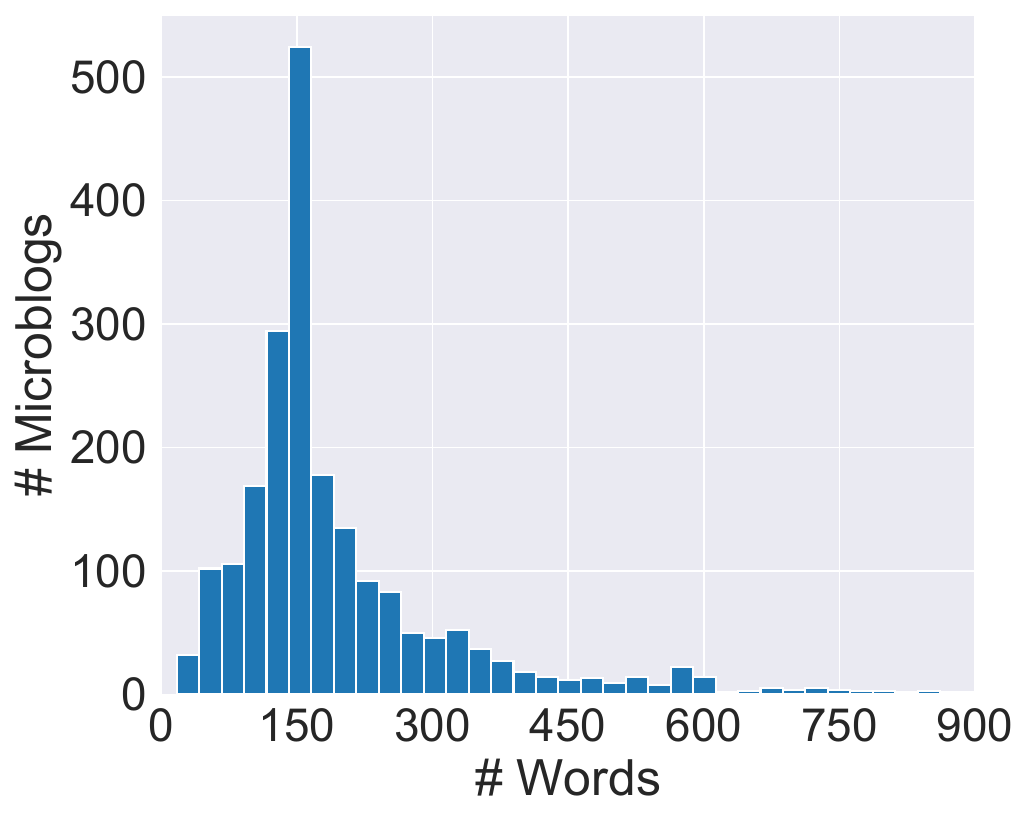}
        \caption{Dist. of Words}
        \label{fig:news_word}
        \end{minipage}~
        \begin{minipage}[b]{0.345\textwidth}
        \includegraphics[width=\textwidth]{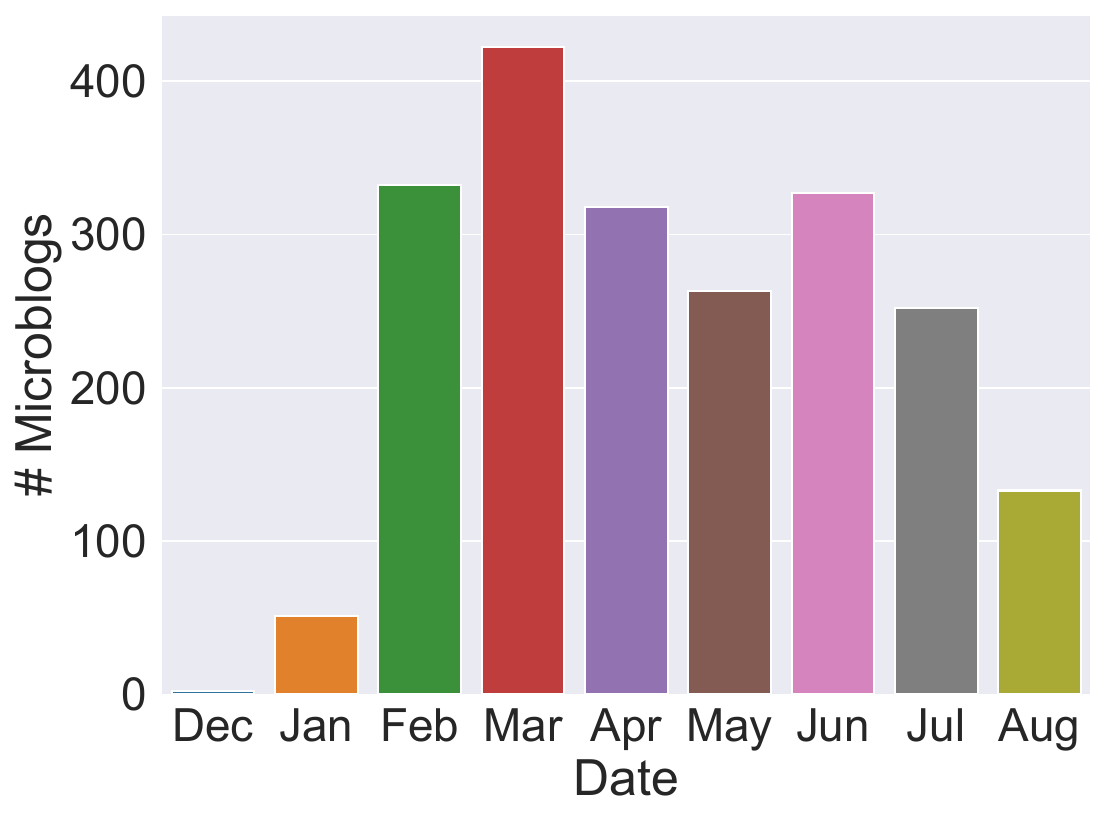}
        \caption{Dist. of Dates Posted}
        \label{fig:news_date}
        \end{minipage}~
        \begin{minipage}[b]{0.31\textwidth}
        \includegraphics[width=\textwidth]{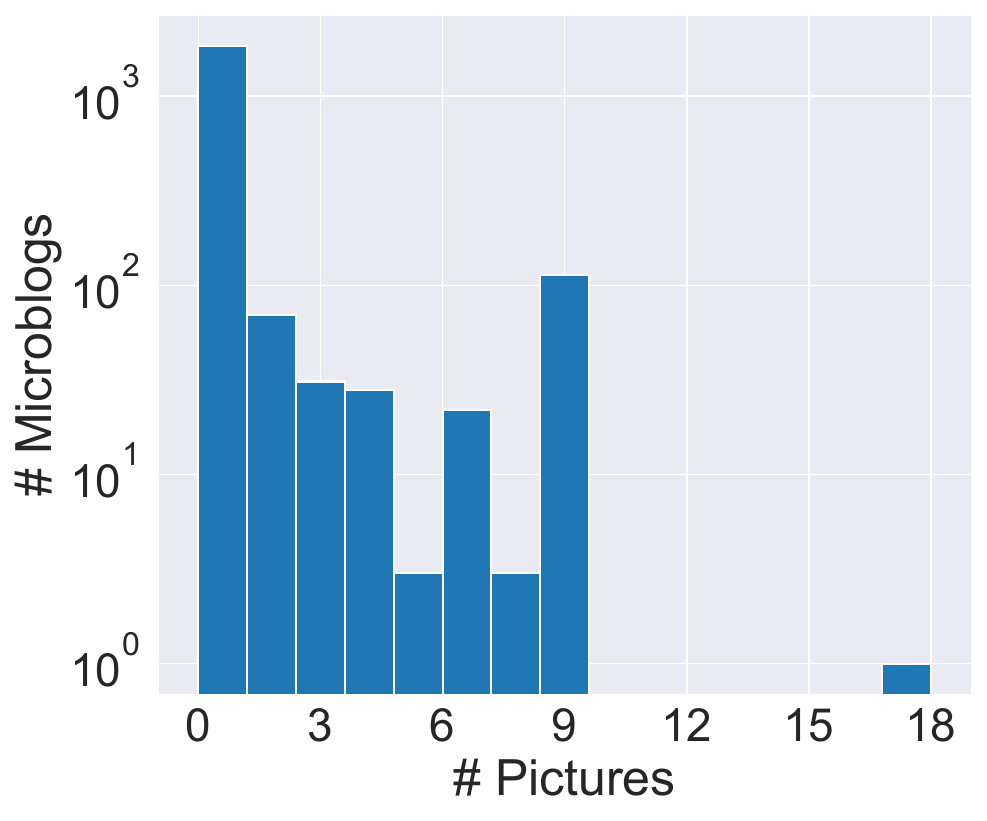}
        \caption{Dist. of Images}
        \label{fig:news_image}
        \end{minipage}
    \end{minipage}
\end{figure}
\begin{figure}[t]
    \begin{minipage}{\textwidth}
        \begin{minipage}[b]{0.325\textwidth}
        \includegraphics[width=\textwidth]{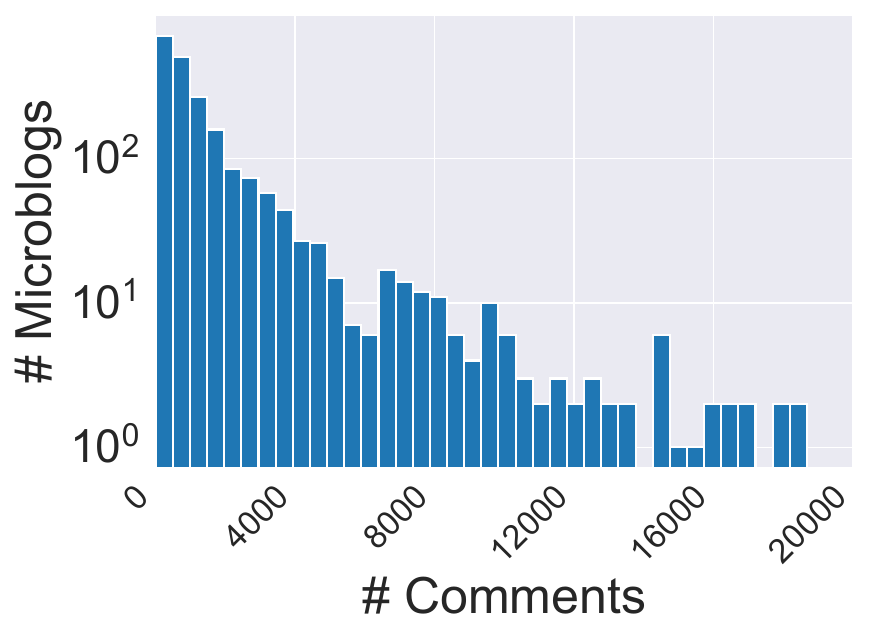}
        \caption{Dist. of Comments}
        \label{fig:news_comments}
        \end{minipage}~
        \begin{minipage}[b]{0.325\textwidth}
        \includegraphics[width=\textwidth]{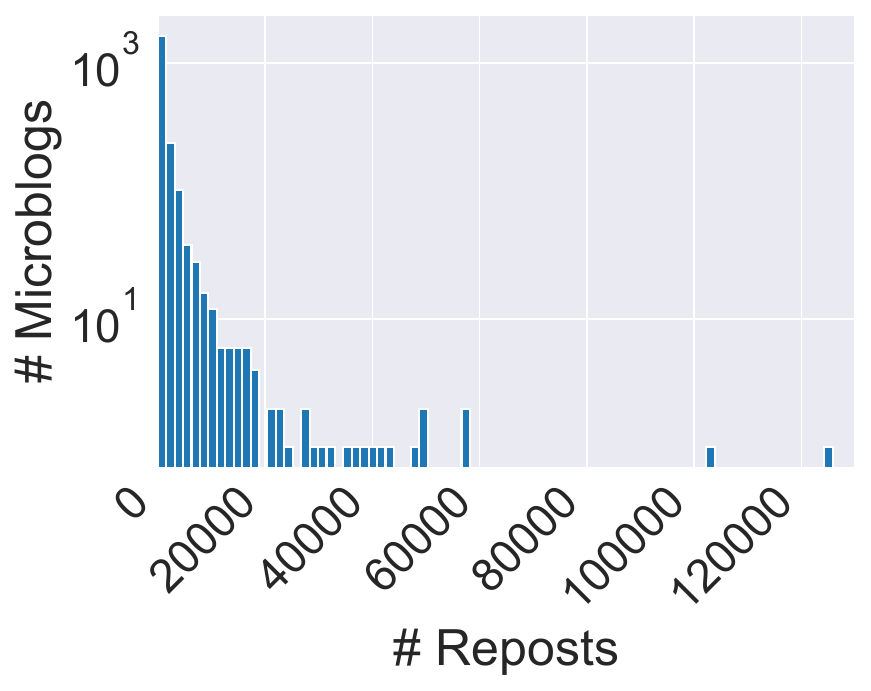}
        \caption{Dist. of Reposts}
        \label{fig:news_forwards}
        \end{minipage}~
        \begin{minipage}[b]{0.325\textwidth}
        \includegraphics[width=\textwidth]{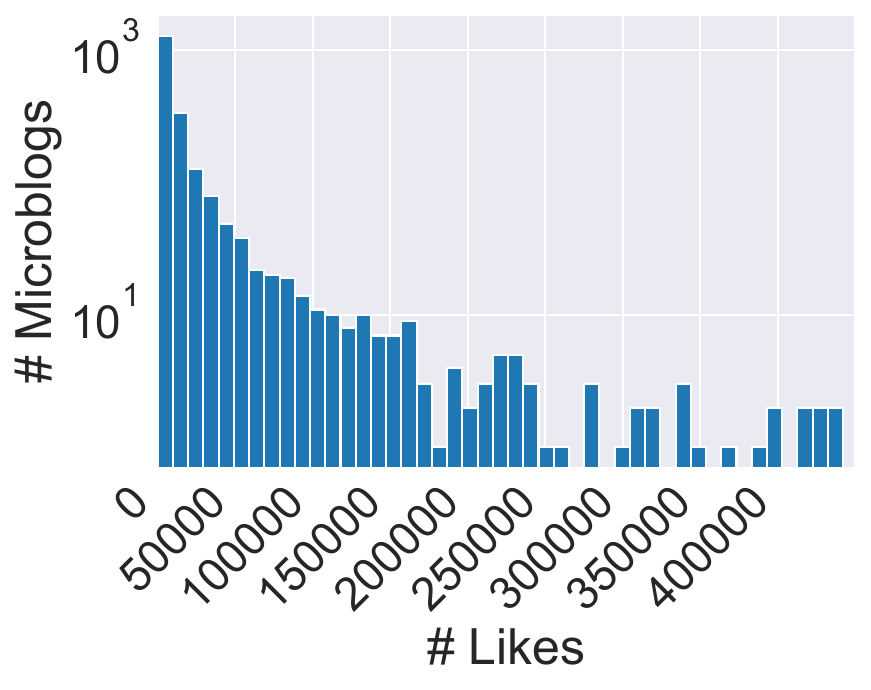}
        \caption{Dist. of Likes}
        \label{fig:news_likes}
        \end{minipage}
    \end{minipage}
\end{figure}

\paragraph{I. Textual Information of Microblogs}

\begin{CJK*}{UTF8}{gbsn}
We first present the distribution of our keywords within the collected microblogs in Figure \ref{fig:keyword_dist}; the keywords help identify if a microblog is relevant to COVID-19. We observe that the Chinese keywords are more frequently used than the English ones; this is reasonable as Weibo mostly targets Chinese users who might be multilingual.

On the other hand, Figure \ref{fig:word_cloud} presents the word cloud of collected microblogs for all words in addition to the keywords. The top-ten vocabularies with the highest frequencies are ``病例 (case)'' (\#=2919), ``确诊 (confirmed case)'' (\#=2158), ``新冠 ((abbr.) COVID-19)'' (\#=1827), ``疫情 (pandemic/epidemic)'' (\#=1701), ``新冠肺炎 (COVID-19)'' (\#=1649), ``新增 (new case)'' (\#=1331), ``美国 (United State)'' (\#=1227), ``检测 (test)'' (\#=923), and ``中国 (China)'' (\#=866). We observe that most of these vocabularies are included in our keyword list, while a few are not. These are not directly related to COVID-19, e.g., United States and China.
\end{CJK*}

Finally, in Figure \ref{fig:news_word}, we plot the distribution of the number of words within collected microblogs, which share an average (median) number of 216 (154).
We point out that each microblog is restricted to up to 140 words for all Weibo users before version 6.0.0, after which the restriction has been lifted for VIP users, e.g., People's Daily; it explains why some microblogs in our data have over 140 words.

\paragraph{II. Temporal and Visual Information of Microblogs} 

Figure \ref{fig:news_date} presents the distribution of the posting time of our collected microblogs. We observe that a very few microblogs related to COVID-19 were posted in December 2019 as the first case of the coronavirus was confirmed at the end of this month~\cite{sohrabi2020world}. We notice that the number of COVID-19 microblogs (or say, online discussion) significantly increases since February 2020, which is also the month with the highest number of confirmed cases of the coronavirus.\textsuperscript{\ref{ft:who}}

As for images attached within microblogs, we observe from Figure \ref{fig:news_image} that most collected microblogs contain no more than 9 images.
A few microblogs have more than nine images because posting over 9 images within a microblog has been allowed recently, as a new version 9.9.3 of Weibo.

\paragraph{III. Propagation and Responses of Microblogs} We track the influence of [real and fake] microblogs on the coronavirus, as well as how they propagate on Weibo from three perspectives: (i) comments, (ii) forwards, and (iii) likes; their corresponding distributions for our data are provided in Figures \ref{fig:news_comments}-\ref{fig:news_likes}, which all follow a power-law-like distribution with a long tail.
We observe that (i) in general, $\sim$80\% microblogs have less than 2,000 comments, 2,000 forwards, and/or 20,000 likes; (ii) the most influential microblog can have over 20,000 comments, one million forwards, and/or one million likes; and (iii) all collected microblogs share an average (median) frequencies of being commented, forwarded, and liked of 1,785 (744), 3,482 (586), and 26,851 (6,179).

\vspace{-2mm}
\section{Benchmark Results with \textsf{CHECKED} data for Fake News Detection}
\label{sec:baseline}

In this section, we provide the benchmark result using \textsf{CHECKED} data when utilizing various methods to predict fake news. These neural-network-based methods have been well-established and widely-accepted for text-classification, including 
(i) FastText~\cite{joulin2017bag}, 
(ii) TextCNN~\cite{kim2014convolutional},
(iii) TextRNN~\cite{liu2016recurrent}, 
(iv) Attention-based TextRNN (Att-TextRNN)~\cite{zhou2016attention}, and the (v) Transformer~\cite{vaswani2017attention}. We first separate the data, based on their posting time order, for training, validation, and testing with a proportion of 70\%:10\%:20\%. Due to the class imbalance in data, we use macro $F_1$ to evaluate the method performance. For all baseline methods, we set the padding size as 150, batch size as 16, and dropout ratio as 0.5. Final results are provided in Table~\ref{tab:basel}. We observe that TextCNN performs best in this scenario, achieving an around 0.938 macro $F_1$ score.

\begin{table}[t]
\centering
\caption{Benchmark Results using \textsf{CHECKED} data to Detect Fake News}
\label{tab:basel}
\begin{tabular}{@{}llllll@{}}
\toprule[1pt]
 & \textbf{FastText} & \textbf{TextCNN} & \textbf{TextRNN} & \textbf{Att-TextRNN} & \textbf{Transformer} \\ 
\midrule[0.5pt]
\textbf{Macro $F_1$} & 0.839 & 0.938 & 0.700 & 0.871 & 0.927 \\ 
\bottomrule[1pt]
\end{tabular}
\end{table}

\vspace{-2mm}
\section{Conclusion}
\label{sec:conclusion}

We present \textsf{CHECKED}, a Chinese COVID-19 dataset containing fact-checked microblogs from Weibo. The dataset includes a total of 2,104 microblogs from December 2019 to August 2020. Each microblog contains ground-truth labels, textual, visual, and propagation information that include 1,868,175 reposts, 1,185,702 comments, and 56,852,726 likes by 737,751 users. We emphasize that \textsf{CHECKED} can be only used for academic research. IDs of microblogs and users have been hashed in the dataset to protect users' privacy. The dataset can be improved by (i) involving more COVID-19 microblogs from other sources of real news; (ii) including the data on COVID-19 news articles from other websites spreading on Weibo. We hope this dataset can promote research on COVID-19 fake news. 


%
%

\vspace{-2mm}
\bibliographystyle{spmpsci}      
\bibliography{myRefs-background,myRefs-fake-review,myRefs-fate,myRefs-knowledge,myRefs-others,myRefs-own-work,myRefs-propagation,myRefs-resources,myRefs-sources,myRefs-style,newRef}   

\end{document}